\documentclass[12pt]{article}
\usepackage{epsf,epsfig,here}
\usepackage{graphics}
\usepackage{graphicx}
\usepackage{epic}
\usepackage{floatflt}
\usepackage{dcolumn}
\usepackage{amsmath}
\RequirePackage{xspace}

\begin{document}
\def\invfb   {\ensuremath{\mbox{\,fb}^{-1}}\xspace}
\def\invab   {\ensuremath{\mbox{\,ab}^{-1}}\xspace}
\def\twobg  {\ensuremath{2 \beta + \gamma}\xspace}
\def\Kz    {\ensuremath{K^0}\xspace}
\def\Bz    {\ensuremath{B^0}\xspace}
\def\Btag    {\ensuremath{B_{\rm tag}}\xspace}
\def\Bbar  {\kern 0.18em\overline{\kern -0.18em B}{}\xspace}
\def\Bb    {\ensuremath{\Bbar}\xspace}
\def\Bzb   {\ensuremath{\Bbar^0}\xspace}
\def\BzBzb {\ensuremath{B^0 -{\kern -0.16em \Bzb}}\xspace}
\def\Bzb   {\ensuremath{\Bbar^0}\xspace}
\def\BtoDKzPi{\ensuremath{B^0 \to D^- K^0 \pi^+}}
\def\Dmd{\ensuremath{{\rm \Delta}m_d}\xspace}

\newcommand{\jprl}      [1]  {{Phys.\ Rev.\ Lett.\ {\bf #1}}} 
\newcommand{\jprd}      [1]  {{Phys.\ Rev.\ D~{\bf #1}}}

  \newcommand{\rhobar}{\bar {\rho}}
  \newcommand{\etabar}{\bar{\eta}}
  \newcommand{\epsilonk}{\varepsilon_K}
  \newcommand{\vubovcb}{\left | \frac{V_{ub}}{V_{cb}} \right |}
  \newcommand{\vubsvcb}{\left | V_{ub}/V_{cb}  \right |}
  \newcommand{\vtdovts}{\left | \frac{V_{td}}{V_{ts}} \right |}
  \newcommand{\epsp}{\frac{\varepsilon^{'}}{\varepsilon}}
  \newcommand{\dms}{\Delta m_s}
  \newcommand{\snb}{sin 2\beta}
  \newcommand{\BK}{B_K}
  \newcommand{\fbdsqbd}{F_{B_d} \sqrt{\hat B_{B_d}}}
  \newcommand{\fbssqbs}{F_{B_s} \sqrt{\hat B_{B_s}}}
  \newcommand{\vcb}{\left | {V_{cb}} \right |}                      
  \newcommand{\ra}{\rightarrow}

\renewcommand{\arraystretch}{1.2}
\pagestyle{empty}
\pagenumbering{arabic}
\vskip  1.5 cm
\begin{center}
{\Large {\bf Feasibility study for a model independent measurement of \twobg in \\ 
\Bz decays using $D^- K^0 \pi^+$ final states.}}
\end{center}
\vskip 1.0truecm
\begin{center}
{\bf \large F.~Polci$^{(a),(b)}$, M.-H. Schune$^{(b)}$ and A. Stocchi$^{(b)}$}
\end{center}
\vskip 0.3truecm
\begin{center}
{\small
\noindent
{\bf $^{(a)}$ Dipartimento di Fisica, Universit\`a di Roma ``La Sapienza'' \\ and INFN, Sezione di Roma,}\\
\hspace*{0.5cm}{Piazzale A. Moro 2, 00185 Roma, Italy}\\
\noindent
{\bf $^{(b)}$ Laboratoire de l'Acc\'el\'erateur Lin\'eaire,}\\
\hspace*{0.5cm}{IN2P3-CNRS et Universit\'e de Paris-Sud, BP 34, 
F-91898 Orsay Cedex}\\
}
\end{center}
\vspace*{0.5cm}

\begin{abstract}
In this paper we present a feasibility study for measuring the combination of the Unitarity Triangle angle 
2$\beta +\gamma$ with a time dependent Dalitz analysis in \Bz decays using $D^- K^0 \pi^+$ final states following 
the method proposed in \cite{aps}.
For this study we build a model for this decay using the latest experimental information and we 
investigate the possibility of fitting together with 2$\beta +\gamma$ all the relevant strong 
amplitudes and phases of the decay model.
\end{abstract}

\section{Introduction}

The interference effects between the $b \to c$ and $b \to u$ decay amplitudes in the 
time-dependent asymmetries of $B$ decaying into $D^{(*)}\pi$ and $D^{(*)}\rho$ final 
states allow for the determination of $2\beta$+$\gamma$. Nevertheless the extraction of $2\beta$+$\gamma$
depends upon the knowledge of the absolute value of the ratio ($r$) of the amplitudes 
( e.g.: $r=|{\cal A}(\overline{B}^0 \to D^- \pi^+)| / |{\cal A}(B^0 \to D^-\pi^+)|$) 
and of the strong phase difference, which nave not been determined experimentally so far. 
Furthermore, the ratio $r$ is expected to be rather small, 
of the order of $\lambda |V_{ub}/V_{cb}| \simeq$ 0.02. In addition the extraction of 2$\beta +\gamma$ 
from the two-body decays suffers from a eight-fold ambiguity.
One can estimate the ratio $r$ by deducing for instance the branching fraction 
$Br(B^0 \rightarrow D^{(*)-} \pi^+)$ from the measured branching fraction $Br(B^0 \rightarrow D_s^{(*)-}\pi^+)$. 
This approach is valid up to a theoretical uncertainty 
related to SU(3) breaking effects and to the size of annihilation processes which do not contribute to
$D_s^{(*)-}\pi^+$ final states.
Using all the experimental constraints~\cite{PDG},\cite{Aubert:2005iq}, a tentative extraction 
of 2$\beta +\gamma$ has been made, attributing 100$\%$ uncertainty to these assumptions, and 
getting \cite{ref:utfit} :  2$\beta +\gamma = (\pm 90 \pm 46)^o$ (the $\pm$ in front of the central
value refers to the fact that 2$\beta +\gamma$ is determined up to a $\pi$ ambiguity).

Another way of measuring  $\twobg$ is to perform a time dependent Dalitz analysis of the
three-body decay : \BtoDKzPi \footnote {Charge conjugation is implied throughout the paper unless otherwise stated}, 
as proposed in \cite{aps}. This method allows 
in principle to overcome the disadvantages encountered in the determination of 2$\beta +\gamma$ from 
$D^{(*)}\pi$ or $D^{(*)}\rho$ decays.
First of all, in the region of the Dalitz plane where most of the interference takes place the value of the 
ratio $r$ is expected to be of the order of 0.4 since it involves $b \to c$ and $b \to u$ colour suppressed diagrams. 
Furthermore $\twobg$ can be measured with only a two-fold ambiguity as explained in \cite{ref:ap}.
In addition, this method could be theoretically clean, because the strong amplitudes and phases 
can be, in principle, measured from data. 

Early  studies can be found in \cite{ref:ap}. In this paper we present a feasibility study 
of the analysis under realistic conditions. Unlike to previous studies, we have used a model for the Dalitz 
structure of the \BtoDKzPi  decays that uses the latest experimental 
measurements. We have also included realistic $\Bz \-- \Bzb $
tagging performances and explored the possibility of fitting the strong 
amplitudes and phases of the most relevant interfering resonances by using both the tagged and untagged 
events. Finally we have studied the effect of background in the determination of $\twobg$.

\section{The \BtoDKzPi decays}
\subsection {Time dependence : general case}
The measurement of $\twobg$ can be achieved through a study of the time-dependent 
evolution of \Bz decay sensitive to that phase.  This can only be performed on tagged 
events, that means events for which the original flavour of the reconstructed $B$ is known. In a $B$ factory this 
is possible through the determination of the flavour of the other $B$ in the event 
since the two $B$ mesons are produced in a coherent $J^{PC}=1^{--}$ state. 
Thus if we determine that at a time $t$ we have a \Bz, then the other $B$ in the event 
at the same time $t$ is a \Bzb . In practice, one of the $B$ (the \Btag) is reconstructed 
in a flavour tagging state. The time-dependent decay rates taking into account 
CP violation for a \Bz ($P_{+}$) or \Bzb ($P_{-}$) tagging meson are given by :  
\begin{equation} 
P_{\pm} 
        = \frac{N}{4 \tau} e^{- t/\tau} ( 1 \mp C \cos (\Dmd t) \pm S \sin (\Dmd t ) )
\label{eq:fplus}
\end{equation} 
where $\tau$ is the \Bz lifetime, $\Delta m_d$ is the \BzBzb mixing 
frequency and $N$ a normalisation factor. 
The parameter $S$ is non-zero if there is mixing-induced CP violation, while a non-zero 
value for $C$ would indicate direct CP violation. 

\subsection{The \BtoDKzPi case}
\begin{figure}[!tbp]
\begin{center}
\epsfig{figure=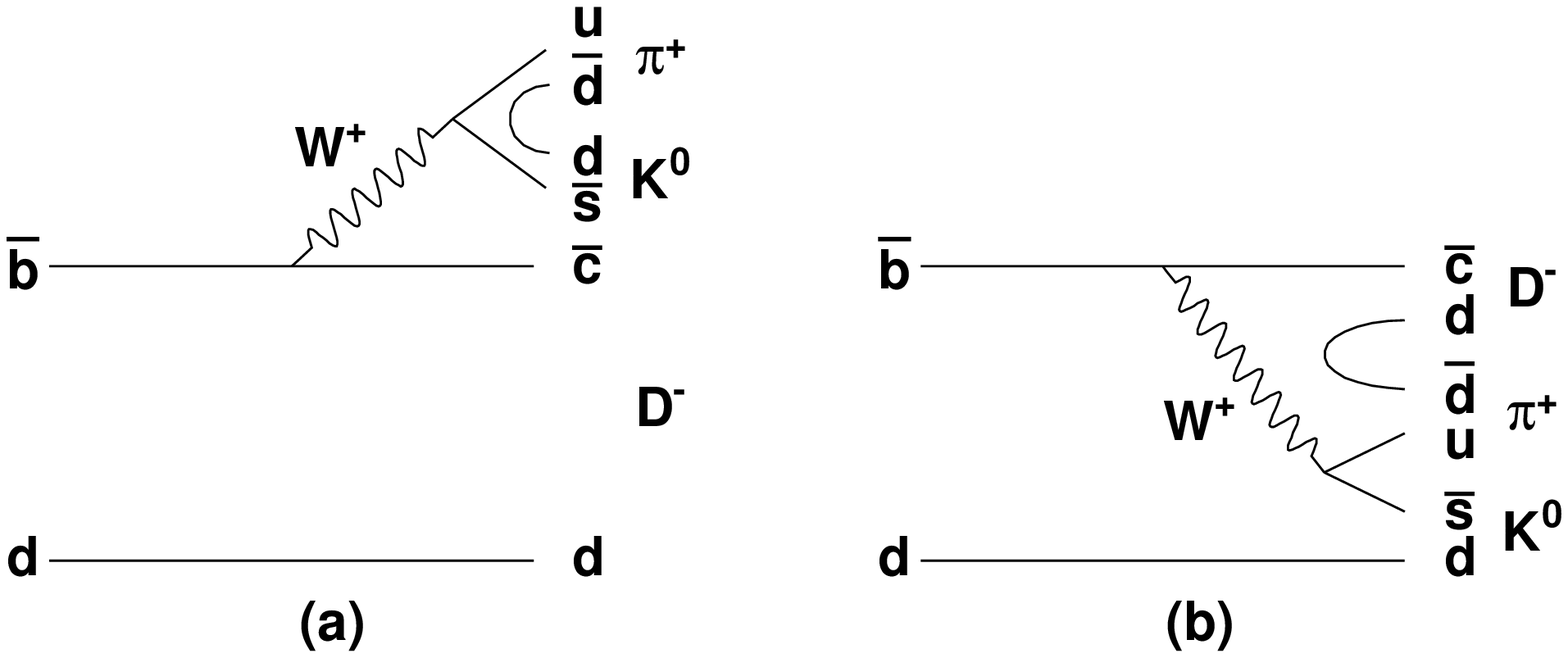,width=9.0cm}
\epsfig{figure=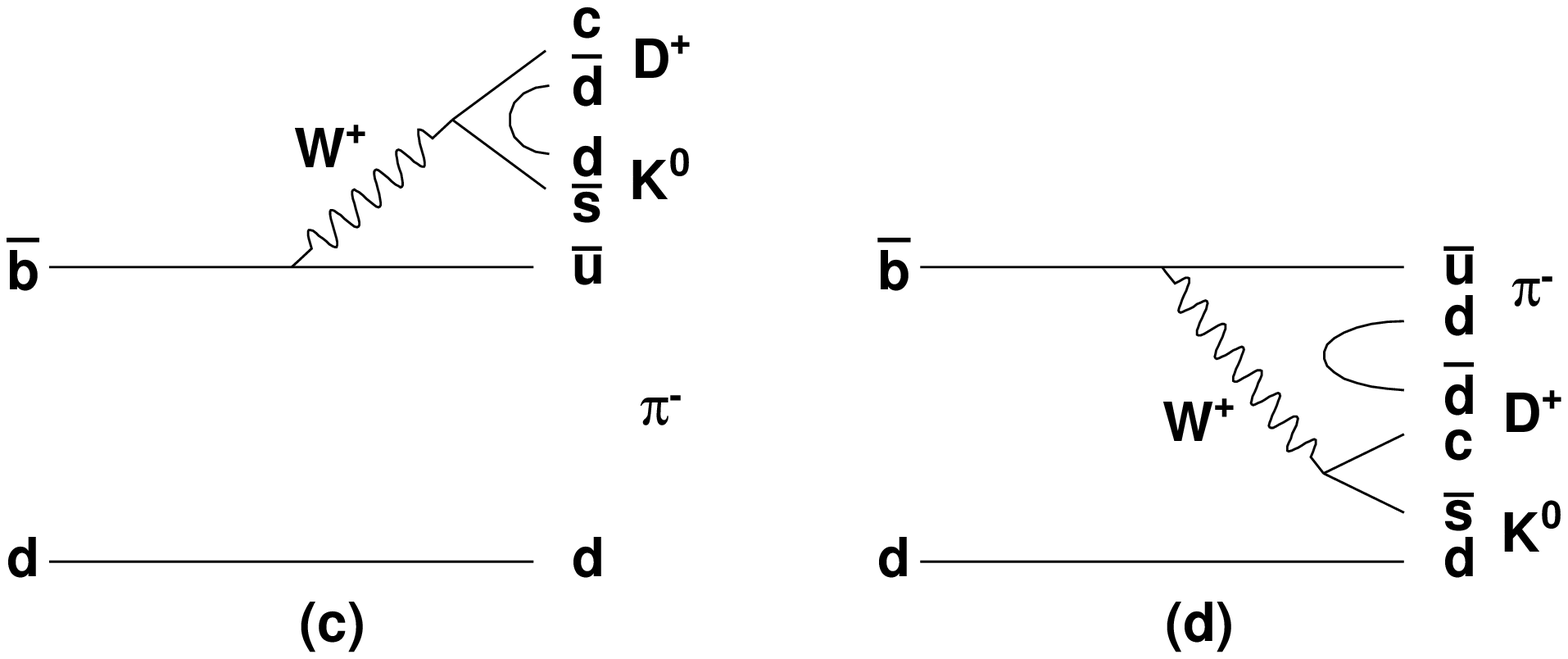,width=9.0cm}
\end{center}
\caption{\it {Feynman diagrams describing the processes contributing to the decays \BtoDKzPi. 
a) $B^0 \rightarrow D^- K^{*+} (K^0 \pi^+)$ and higher $K^{**}$ resonances, 
b) $B^0 \rightarrow \bar{D}^{**0} (D^- \pi^+) K^0$, 
c) $B^0 \rightarrow {D_s}^{**+} (D^+ K^0) \pi^-$, 
d) $B^0 \rightarrow {D}^{**0} (D^+ \pi^-) K^0$. 
The processes in a,b (c,d) are $V_{cb}$ ($V_{ub}$) mediated.}}
\label{DIAGRAMS}
\end{figure}

The final state $D^- K^0 \pi^+$ can be reached through the diagrams shown in Figure \ref{DIAGRAMS} and considering the
\BzBzb mixing.
The parameters $S$ and $C$ of Eq. \ref{eq:fplus} will depend on the position in the $B$ Dalitz plot.
Indeed, a given bin in the Dalitz plot gets contributions from amplitudes and strong phases of $b \rightarrow c$ 
and $b \rightarrow u$ transitions. 

The  model assumed for the decay parametrises the amplitude $A$ at each point $k$ of the 
Dalitz plot as a sum of two-body decay matrix elements and a non-resonant term according
to the following expression :
\begin{equation}
A_{c_k(u_k)}e^{i\delta_{c_k(u_k)}}=\sum_j a_je^{i\delta_j} BW^j_k(m,\Gamma,s) + a_{nr} e^{i \phi_{nr}}
\label{eqres}
\end{equation}

\noindent
where $c_k$ ($u_k$) indicates the Cabibbo allowed (suppressed) decay in each point $k$ of the Dalitz plot. 
Each term of the sum is parametrised with an amplitude ($a_j$ or $a_{nr}$) and a phase ($\delta_j$ or $\phi_{nr}$). 
The factor $BW^j_k(m,\Gamma,s)$ 
gives the Lorentz invariant expression for the matrix element of a resonance $j$ as a function of the position $k$
in the $B$ Dalitz plot; the functional dependence varies with the spin $s$ of the resonance according
to the isobar model~\cite{isobar}.
The total phase and amplitude are arbitrary. We have chosen amplitude unity and phase zero for the mode
$K^{*+}(892)$ decaying into $K^0 \pi^+$.\\
Four transitions have to be considered associated to probabilities of connecting a \Bz or \Bzb 
initial state to a $D^+$ or $D^-$ final states. They are indicated in Table~\ref{tab:sumAmp} .

\begin{table}[htpb]
\begin{center}
\begin{tabular}{|l|l|l|} \hline 
			&  $D^- K^0 \pi^+$ final state   & 	 $D^+ K^0 \pi^-$ final state   \\
\hline
$V_{cb}$ contribution &$< D^- K^0 \pi^+ |T|\Bz> = A_{c_k}e^{i\delta_{c_k}}$&$< D^+ K^0 \pi^- |T|\Bzb> = A_{c_k}e^{i\delta_{c_k}}$ 	\\
\hline
$V_{ub}$ contribution &$< D^- K^0 \pi^+ |T|\Bzb> = A_{u_k}e^{i\delta_{u_k}-i \gamma}$&$< D^+ K^0 \pi^- |T|\Bz> = A_{u_k}e^{i\delta_{u_k}+i \gamma}$ \\ 
\hline
\end{tabular}
\end{center}
\caption{\it {Summary of the amplitudes involved in the \twobg analysis. The index 
$k$ refers to the position in the Dalitz plot. The third column is derived from the second one using CP 
transformation.}} 
\label{tab:sumAmp}
\end{table}
The time dependent evolution of these probabilities can be obtained from the resolution of the 
Schroedinger equation. This leads to a new formulation of Eq. \ref{eq:fplus} in which 
the first argument within the parentheses refers to the production state of the $B$ while the second one indicates
the reconstructed final state\footnote{If one neglects $b \rightarrow u$ transitions, the charges of the particles in the final state tag the $B$ decay flavour and one is left with the standard mixing formulae :
$P(\Bz,\Bz)= P(\Bzb,\Bzb) \propto (1-\cos(\Delta m_d t))$ and 
$P(\Bz,\Bzb)= P(\Bzb,\Bz) \propto (1+\cos(\Delta m_d t))$}.

\begin{eqnarray}
P(\Bz,  D^+ K^0 \pi^-)=\frac{A_{c_k}^2+A_{u_k}^2}{2} e^{-\Gamma_B t} \{1 - C^k \cos(\Delta m_d t) + S^k_{+}  \sin (\Dmd  t )\} \\ \nonumber
P(\Bzb, D^+ K^0 \pi^-)=\frac{A_{c_k}^2+A_{u_k}^2}{2} e^{-\Gamma_B t} \{1 + C^k \cos(\Delta m_d t) - S^k_{+}  \sin (\Dmd  t )\}  \\ \nonumber
P(\Bz,  D^- K^0 \pi^+)=\frac{A_{c_k}^2+A_{u_k}^2}{2} e^{-\Gamma_B t} \{1 + C^k \cos(\Delta m_d t) + S^k_{-}  \sin (\Dmd  t )\}  \\ \nonumber
P(\Bzb, D^- K^0 \pi^+)=\frac{A_{c_k}^2+A_{u_k}^2}{2} e^{-\Gamma_B t} \{1 - C^k \cos(\Delta m_d t) - S^k_{-}  \sin (\Dmd  t )\}  \\ 
\label{timelikeli}
\end{eqnarray}
with :
\begin{equation}
C^k = \frac{A_{c_k}^2-A_{u_k}^2}{A_{c_k}^2+A_{u_k}^2} \ ;
S^k_{+} = \frac{2 Im (A_{c_k}A_{u_k} e^{i( \twobg )+i(\delta_{c_k}-\delta_{u_k})})}{A_{c_k}^2+A_{u_k}^2} \  \ \mathrm{and} \
S^k_{-} = \frac{2 Im (A_{c_k}A_{u_k} e^{i( \twobg )-i(\delta_{c_k}-\delta_{u_k})})}{A_{c_k}^2+A_{u_k}^2}
\end{equation}

Using these relations, and because of the presence of the terms $BW^j_k(m,\Gamma,s)$ which vary over 
the Dalitz plot, we can fit the amplitudes ($a_j$) and the phases ($\delta_j$) of Eq. \ref{eqres}, together 
with $\twobg$ with only a two-fold ambiguity. 

\subsection{The \BtoDKzPi model \label{sec:model}}
\begin{figure}[!tbp]
\begin{center}
\epsfig{figure=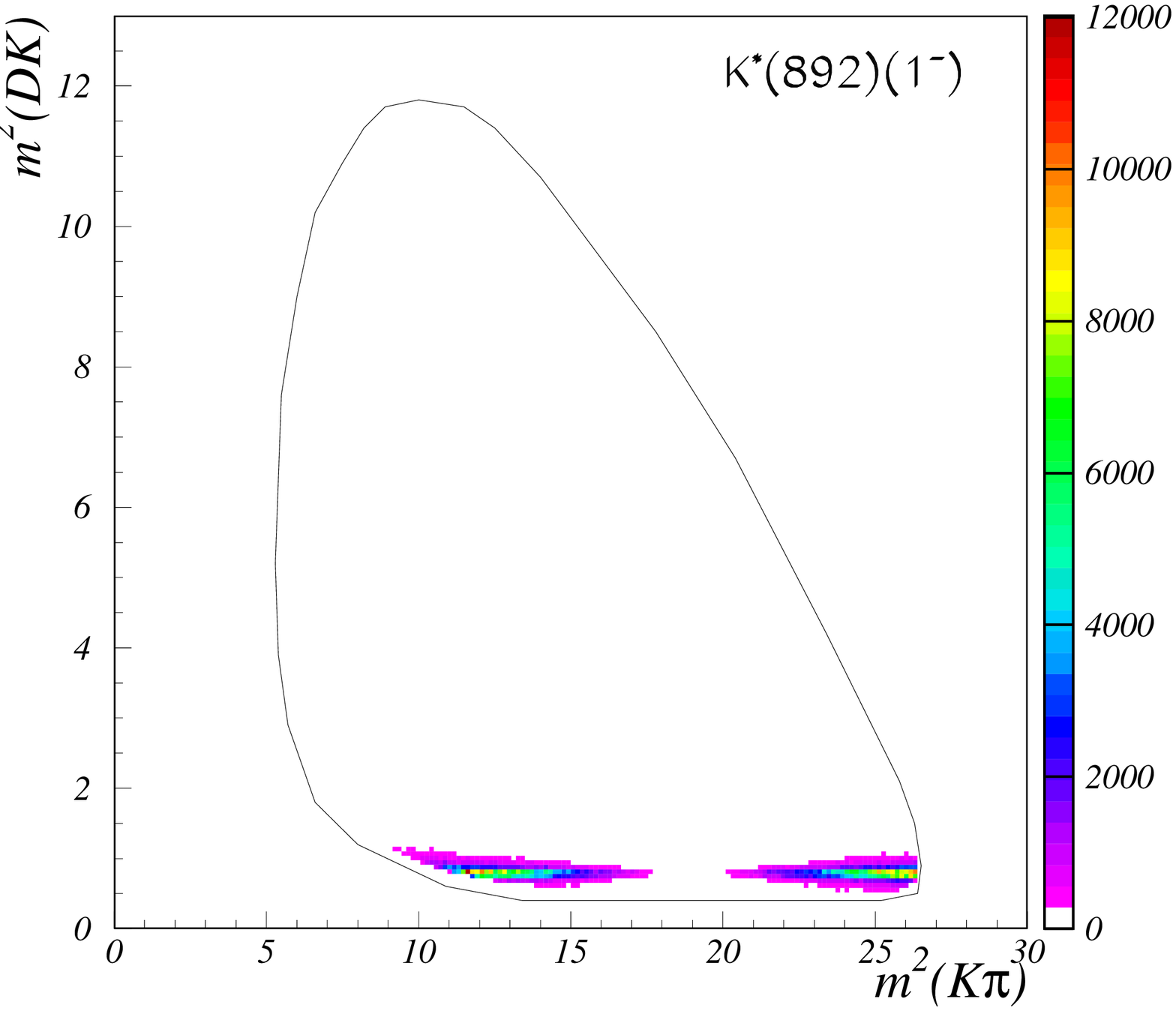,width=5.2cm}
\epsfig{figure=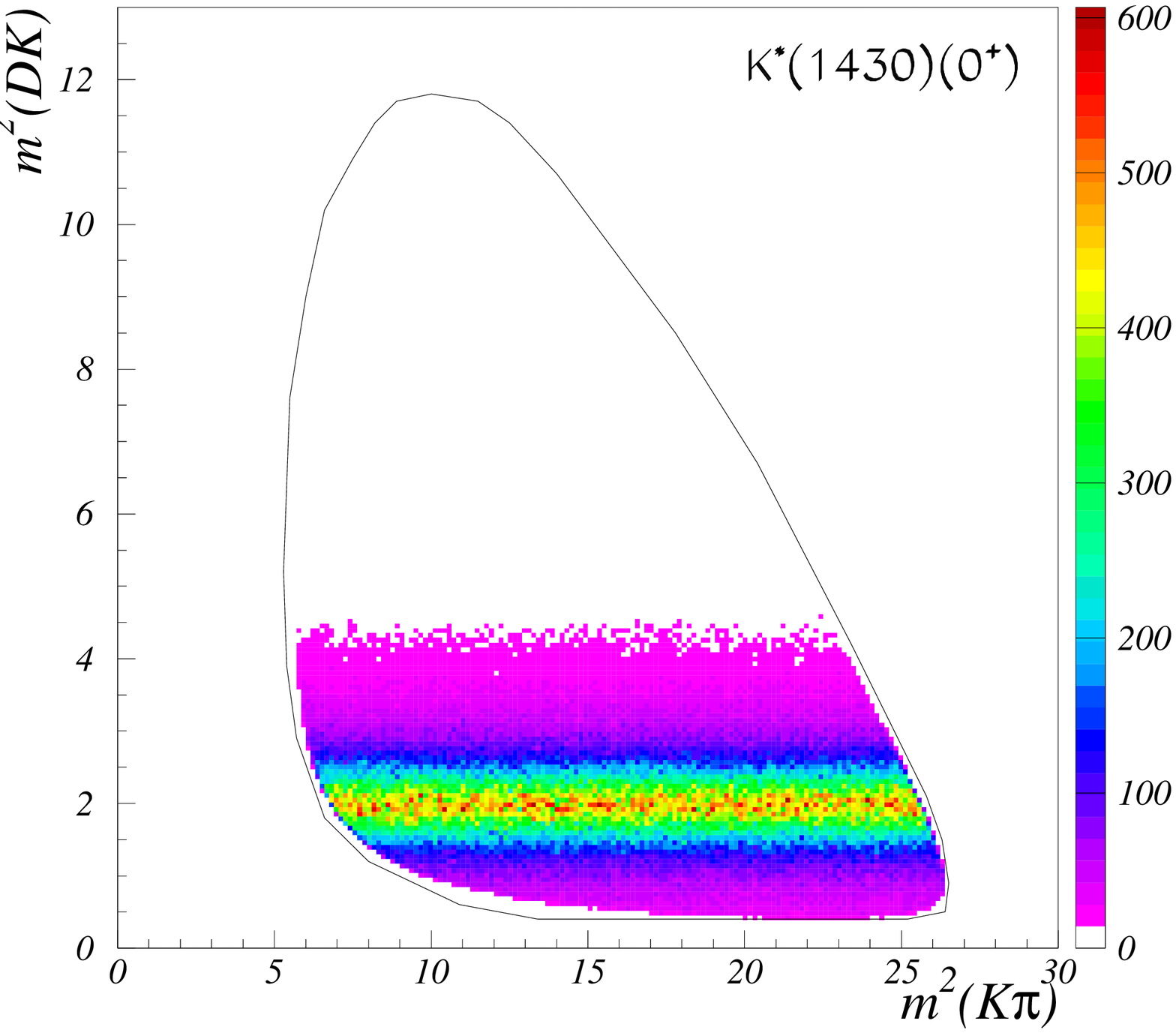,width=5.2cm}
\epsfig{figure=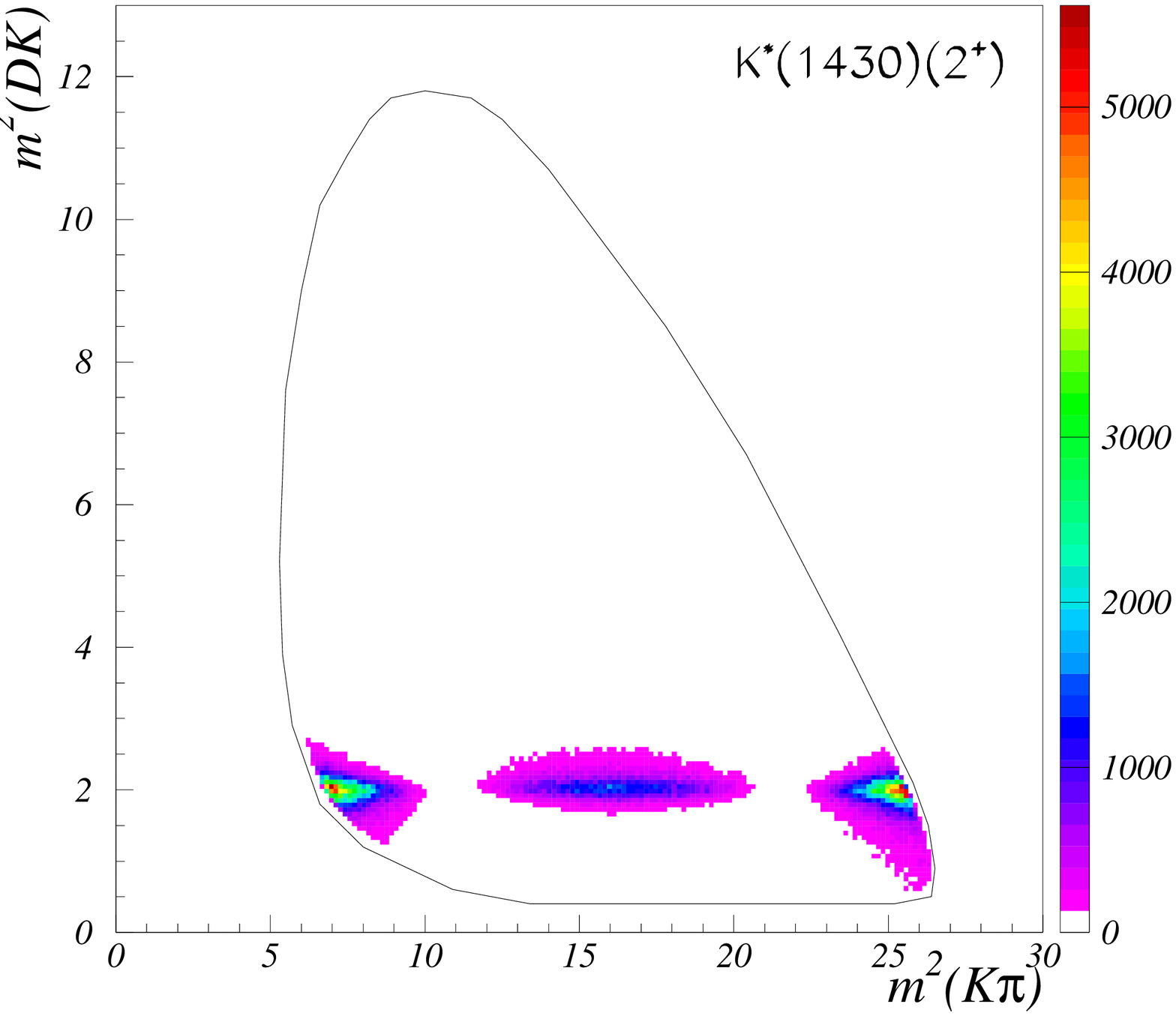,width=5.2cm} \\
\epsfig{figure=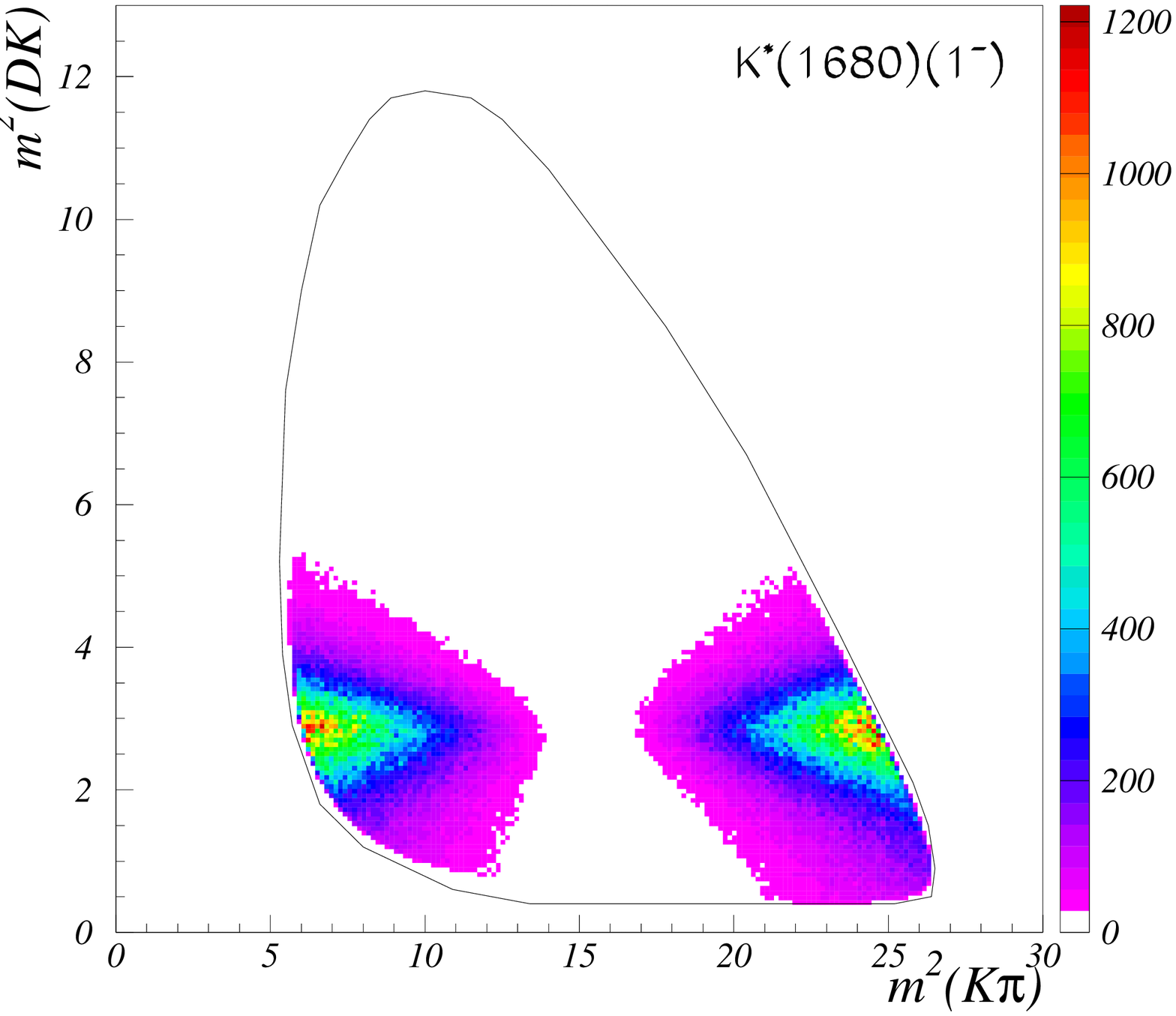,width=5.2cm}
\epsfig{figure=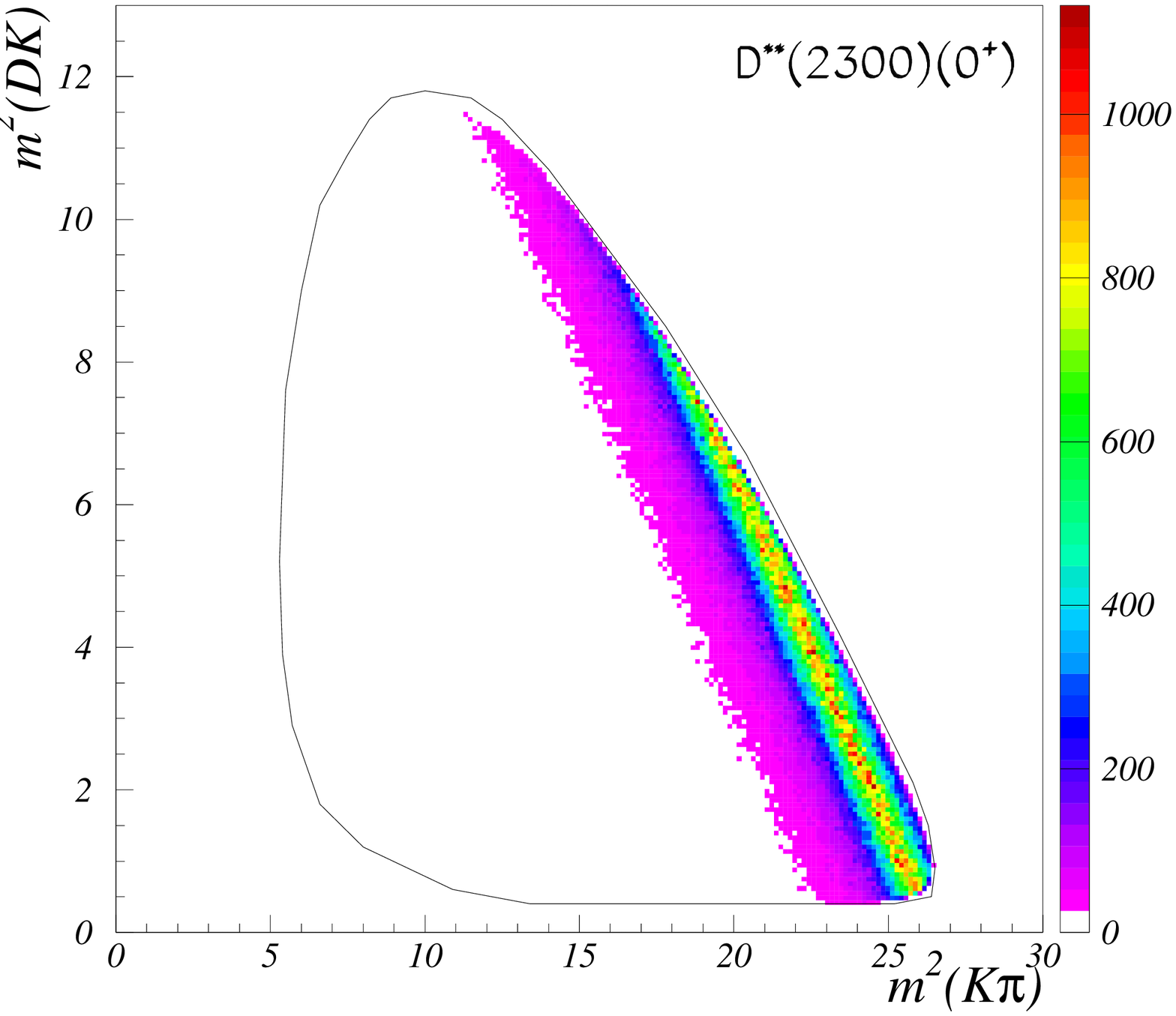,width=5.2cm}
\epsfig{figure=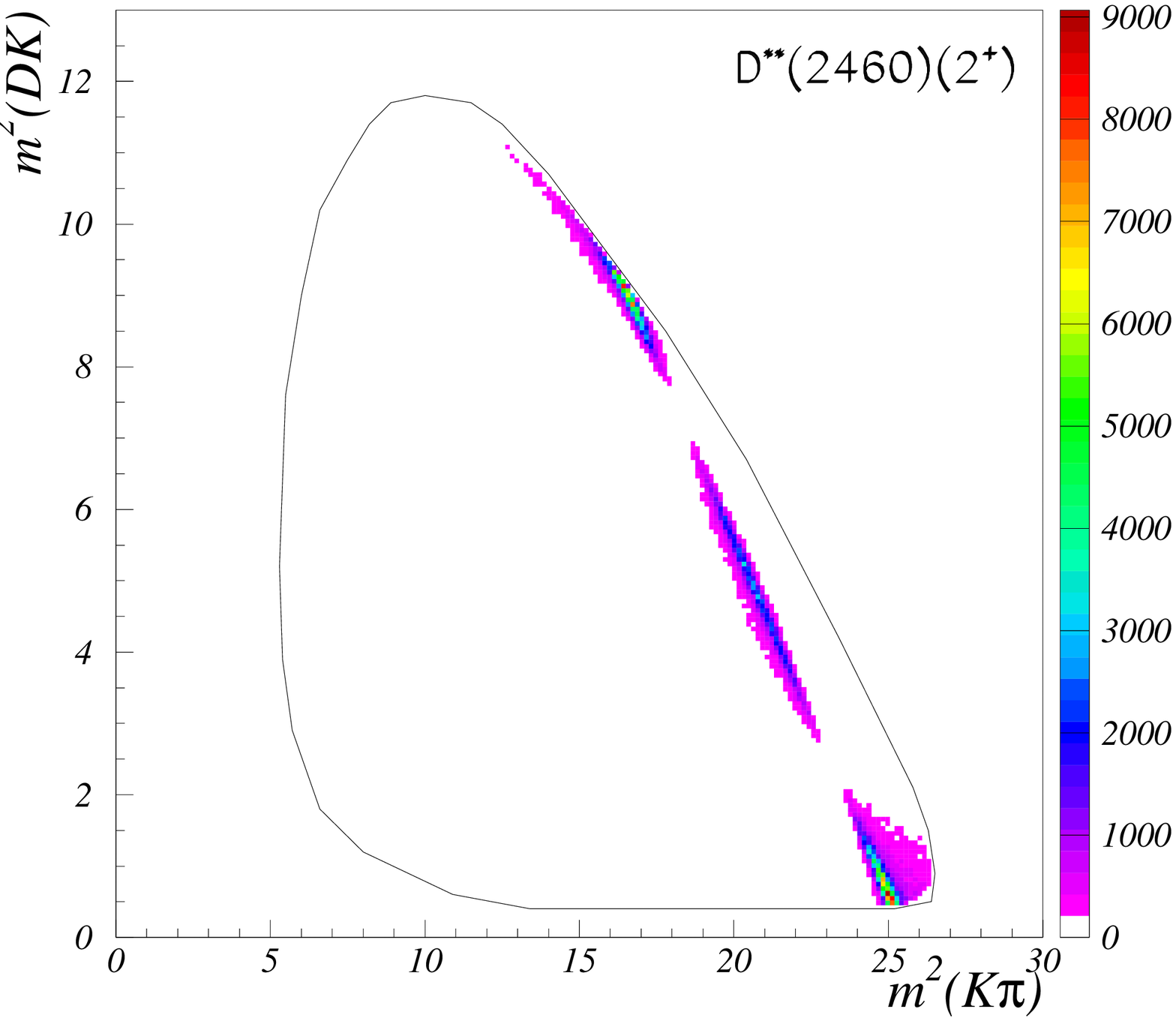,width=5.2cm} \\
\epsfig{figure=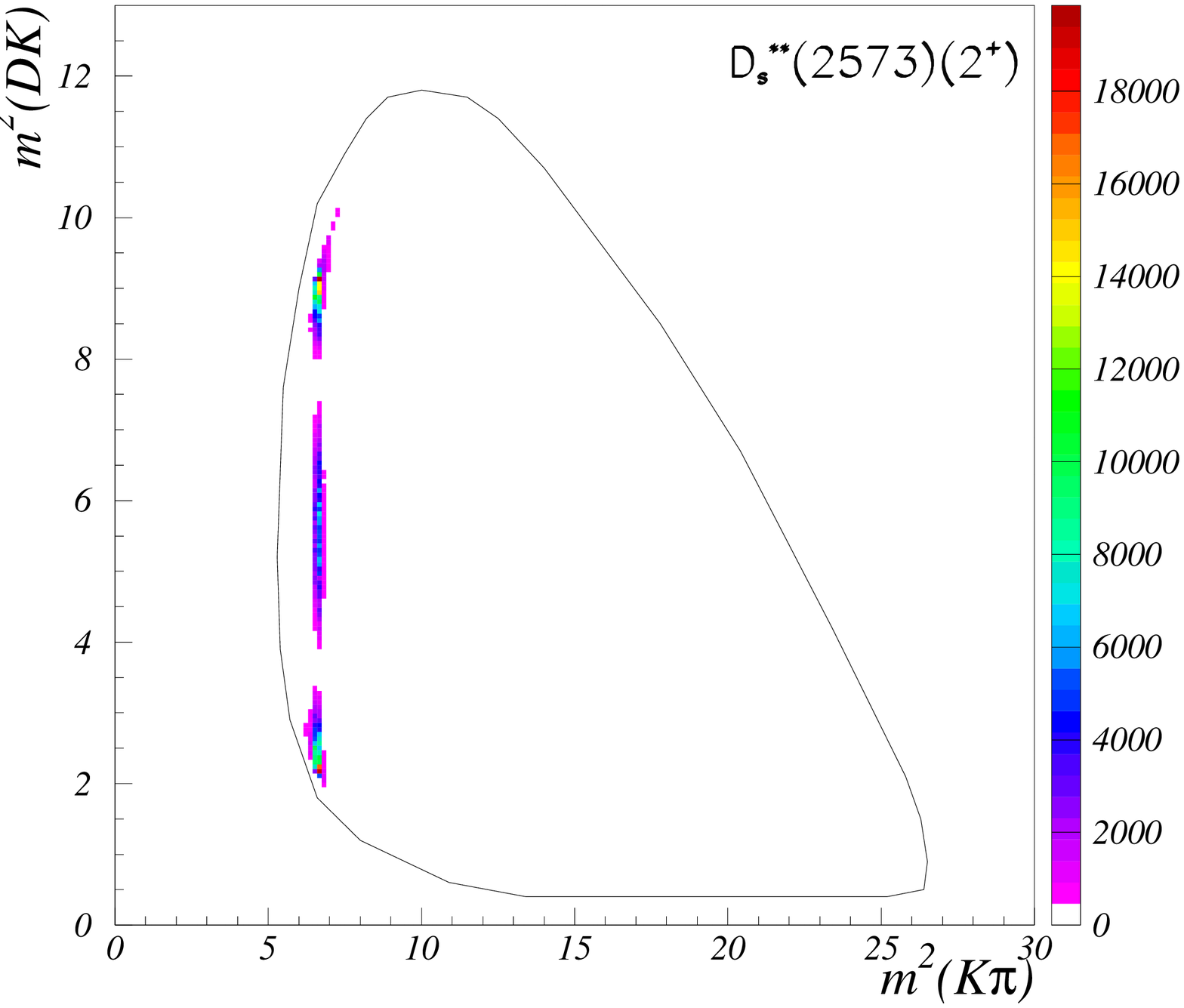,width=5.2cm}
\end{center}
\caption{\it {Various resonances contributing to the  \BtoDKzPi Dalitz plot. From top to bottom, left to right :
$K^{*}(892)^{\pm}, K^{*}_0(1430)^{\pm}, K^{*}_2(1430)^{\pm}, K^{*}(1680)^{\pm}, D_{0}^*(2400)^{0}, D_{2}^*(2460)^{0}$ and $D_{s,2}(2573)^{\pm}$ }}
\label{fig:reson}
\end{figure}

We now discuss the hadronic model for the \BtoDKzPi decay. The resonances taken into account are listed in Table 
\ref{tab:resonances} for both $V_{cb}$ and $V_{ub}$ mediated modes and shown in Figure \ref{fig:reson}.
The Dalitz plot can be modelled in terms of the following resonances :
\begin{eqnarray}
D^-  - X   &  \quad \quad X \ra K^0 \pi^+ \quad & (K^{*}(892)^{\pm}, K^{*}_0(1430)^{\pm}, K^{*}_2(1430)^{\pm}, K^{*}(1680)^{\pm}) \\ \nonumber
K^0  - Y   &  \quad \quad Y \ra D^- \pi^+ \quad & (D_{0}^*(2400)^{0},D_{2}^*(2460)^{0})                                           \\ \nonumber
\pi  - Z   &  \quad \quad Z \ra D^- K^0   \quad & (D_{s,2}(2573)^{\pm}) 
\end{eqnarray}
The $K^0 \pi^+$ resonances ($K^*$ like) can only come from $V_{cb}$ mediated processes, while both $V_{cb}$ and $V_{ub}$ 
transitions contribute to  $D^- \pi^+$ resonances ($D^{**}$ like). Finally the $D^- K^0$ resonances ($D^{**}_s$-like) come 
only through $V_{ub}$ mediated processes.
For the final state under study, ``non resonant'' contributions could come from higher K excited states or 
from higher excited $D^{**}$ states. 
In the first case only $V_{cb}$ processes can contribute while in the second both $V_{cb}$ and $V_{ub}$ processes are contributing. 

The strong phases are not known experimentally and have to 
be chosen arbitrarily. For the values of the amplitudes we can use some 
information available on branching fractions \cite{Aubert:2004at},\cite{PDG}. 
So far we have some information which can help defining the model for the $b \rightarrow c$ part.
The Dalitz plot of this decay has been partially measured in \cite{ref:babarmh}, \cite{ref:babarChinois}.
Two measurements are available to partially define the $D^{**}$ part of the Dalitz plot :
\begin{eqnarray}
\label{dss}
Br(B^0 \rightarrow D^{-} K^{*+})  = (4.6 \pm 0.6 \pm 0.5) \times 10^{-4} \\ \nonumber
Br(B^0 \rightarrow D_{J=2}^{**-} K^+) \times Br(D_{J=2}^{**-} \rightarrow \bar{D^0} \pi^-)  = (1.8 \pm 0.4 \pm 0.3) \times 10^{-5}
\label{eq:dst}
\end{eqnarray}
The  production of the charged $D^{**}$ through the process indicated in the second equation (Eq. \ref{eq:dst}) 
involves only $V_{cb}$ transitions with diagrams at the tree level (T) (similar to the diagram in Figure \ref{DIAGRAMS}-(a), 
with a $D_{J=2}^{**-}$ produced in the lower part and a $K^+$ emitted from the W), while we are interested in the neutral $D^{**}$ 
production which proceed via colour suppressed decays (C) (see Figure \ref{DIAGRAMS}-(b)). For our numerical evaluation of 
the branching fractions of interest for this analysis we have considered that the ratio $|C/T|$ could vary between 0.3 and 0.5. 
Nevertheless we are conscious that it is not straightforward to link production of specific hadronic states mediated by T and C 
suppressed processes.
Furthermore, only $2^+$ final states have been measured so far. We make the hypothesis that the $0^+$ states decaying into D$\pi$ modes 
are as abundant as the $2^+$ states. The possible existence of a non resonant contribution is considered below. \\
For the $b \rightarrow u$ counterpart we use the hypothesis that $r$ is equal to 0.4 (where $r$ in this case is defined 
as $r=\frac{B^0 \rightarrow \bar{D}^{**0} K^0}{B^0 \rightarrow D^{**0} K^0}$). The value of 0.4 is motivated by the fact 
that $r = \frac{|V_{ub}|}{\lambda_{Cab.}|V_{cb}|} \times \frac{{\cal A}(\bar{D}^{**0} K^0)}{{\cal A}(D^{**0} K^0)}$ = 
$\sqrt{\bar{\rho}^2+\bar{\eta}^2} \times \frac{{\cal A}(\bar{D}^{**0} K^0)}{{\cal A}(D^{**0} K^0)}$.
The term $\sqrt{\bar{\rho}^2+\bar{\eta}^2} = 0.408 \pm 0.016 $ \cite{ref:utfit}. The term 
$\frac{{\cal A}(\bar{D}^{**0} K^0)}{{\cal A}(D^{**0} K^0)}$ can be expressed as the ratio of two colour suppressed diagrams 
involving $V_{cb}$ and $V_{ub}$ processes. This ratio is expected to be around unity with an error which is difficult
to estimate. For studying the impact of the $r$ parameter on 2$ \beta +\gamma$ sensitivity we consider a variation for this
parameter in a relatively wide range : between 0.3-0.5.

We also use the measurement of the following ratio \cite{ref:babarmh}:
\begin{equation}
\frac{B^0 \rightarrow D^{-} K^{*+}(892)(K^0 \pi^+)}{B^0 \rightarrow D^{-} K^0 \pi^+} = 0.66 \pm 0.08
\label{ks}
\end{equation}
to distribute the rest of events in the Dalitz plot. It has to be considered that no information on how to distribute these events 
between different excited K states ($K^{*}_0(1430), K^{*}_2(1430), K^{*}(1680)$) is available. 
The contribution from the $D^{**}_s$ resonance in $b \rightarrow u$ transitions is difficult to evaluate. 
By analogy we can consider a similar decay mode : $B^0 \rightarrow D^{*+}D^{*-}K^0$. These decays are mediated by tree diagrams,
however the contributions from $B^0 \rightarrow D^{*\pm} D_{s2}^{\mp}$ final states are small \cite{ref:ddk}.
This result would indicate the predominance of non resonant or very large states. A conservative choice has been 
made not to include them in the present model.
The contributions from the different resonances are fixed at the values given in Table \ref{tab:resonances}. 
Finally we make the conservative hypothesis to have, all over the Dalitz, a small non resonant contribution and in agreement with 
the Standard Model we assume $\twobg = 2$ radians (see for example \cite{ref:utfit}).
All these parameters will be varied to evaluate the impact of the chosen model on the sensitivity of $2\beta + \gamma$. \\
The number of \BtoDKzPi signal events is estimated using the measured 
branching fraction and observed yield~\cite{ref:babarmh}. We have generated 250 signal events per unit of 100 \invfb. 

\begin{table}
\begin{center}
\begin{tabular}{l|c|c|c|c|c|} \hline \hline
                        &$Mass (GeV/c^2)$&$Width (GeV/c^2)$&$J^P$&$a(V_{cb})$ &$a(V_{ub})$ \\ \hline
 $D_{s,2}(2573)^{\pm}$  &    2.572       &      0.015   &   $2^+$ &    -     &  0.02       \\ \hline
 $D_{2}^*(2460)^{0}$    &    2.461       &      0.046   &   $2^+$ &   0.12   &  0.048      \\ \hline
 $D_{0}^*(2308)^{0}$    &    2.308       &      0.276   &   $0^+$ &   0.12   &  0.048      \\ \hline
 $K^*(892)^{\pm}$       &    0.89166     &      0.0508  &   $1^-$ &     1    &  -          \\ \hline
 $K_{0}^*(1430)^{\pm}$  &    1.412       &      0.294   &   $0^+$ &    0.3   &  -          \\ \hline
 $K_{2}^*(1430)^{\pm}$  &    1.4256      &      0.0985  &   $2^+$ &    0.15  &  -          \\ \hline
 $K^*(1680)^{\pm}$      &    1.717       &      0.322   &   $1^-$ &    0.2   &  -          \\ \hline
 Non resonant           &       -        &         -    &    -    &    0.07  &  0.028      \\ \hline \hline
\end{tabular}
\end{center}
\caption{\it List of mass, widths and quantum numbers  of the resonances considered in our model, as 
taken from $PDG2004$. 
The last four columns present the chosen values  of the coefficients $a_j$ and $\delta_j$ in 
Eq.~\ref{eqres} for the Cabibbo allowed and Cabibbo suppressed decays respectively. Note that the choices
for the phases are arbitrary and are not indicated in the Table. In the numerical analysis we have evaluated
the effect of a different choice of strong phases.}
\label{tab:resonances}
\end{table}

\subsection{Sensitivity study for \BtoDKzPi decays}

\begin{figure}[btph]
\begin{center}
\epsfig{figure=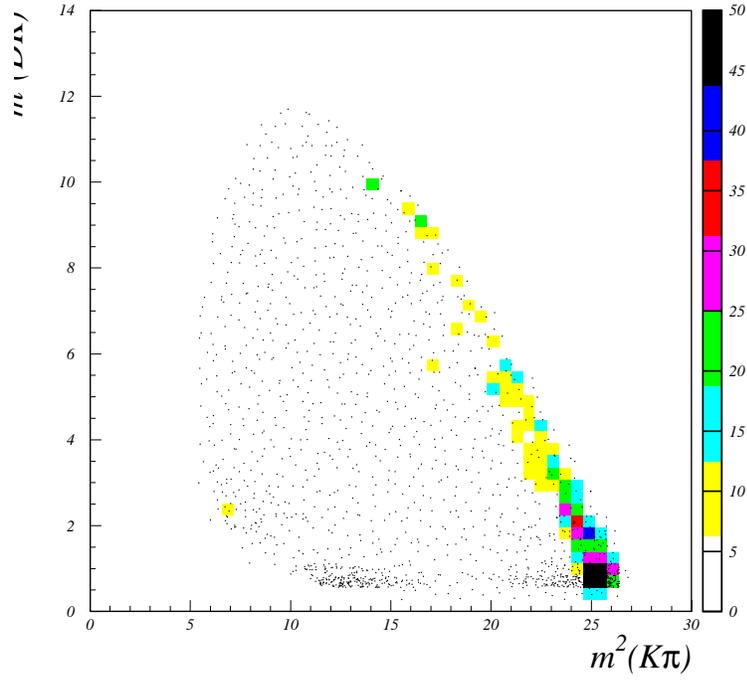,width=10cm}  
\epsfig{figure=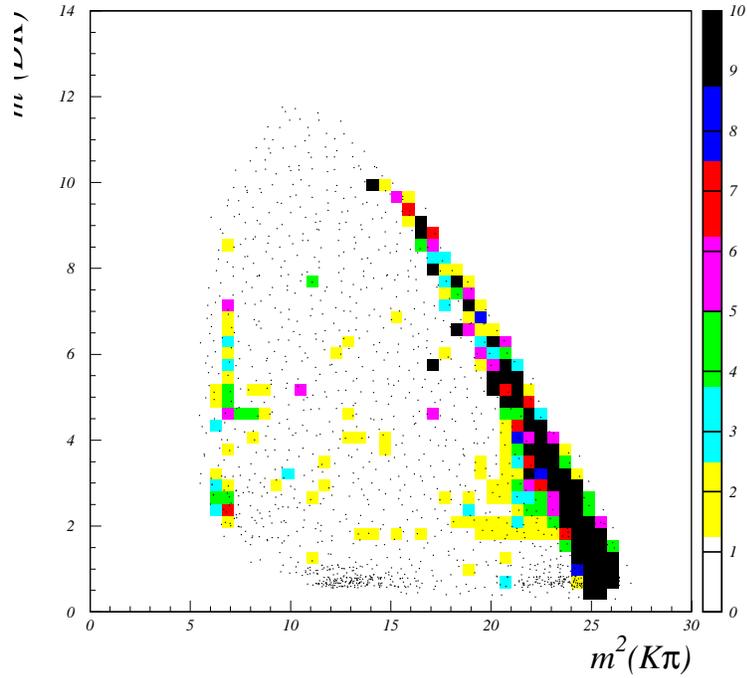,width=10cm} 
\caption{\it { a) Dalitz distribution of the very high statistics Monte-Carlo
sample of signal events. Each event is entering into the plot with a weight given
by the value of the second derivative with respect to 2$\beta + \gamma$ of the
log-likelihood. The black points correspond to the same events with weight equal to unity. 
b) Distribution similar to Figure-a except that we have fixed a maximum value for the weight to be plotted
in order to see in a finer way the structure of the weights over the Dalitz.}}
\label{sensitivity}
\end{center}
\end{figure}

In order to show which are the regions in the Dalitz plot that contribute most in the determination of 2$ \beta + \gamma$, 
we have evaluated, on an event by event basis, the second derivative with respect to 2$\beta + \gamma$ of the log-likelihood  
$\frac{\partial ^2 logL}{\partial ^2 (\twobg})$ constructed according to Eq.\ref{timelikeli} (this likelihood is signal only and 
does not consider resolution effects). A very high statistics Monte-Carlo sample of signal events has been generated 
according to the nominal model described in Section \ref{sec:model}.
The result is shown in Figure \ref{sensitivity} where each event is weighted by the value of the second derivative 
with respect to $\twobg$ of the log-likelihood:
\begin{equation}
weight=\frac{\partial ^2 ln L}{\partial^2 (\twobg)}
\end{equation}
The choice of this weight is motivated by the fact that the uncertainty on $\twobg$ is equal to 
$\sqrt{\frac{1}{\sum{\frac{\partial ^2 ln L}{\partial ^2(\twobg)}}} }$. 
Figure \ref{sensitivity} shows the distribution of the Monte Carlo events in the   $M^2_{D^+ K^0}$ versus $M^2_{K^0\pi^-}$ 
 plane (black dots) and the corresponding sensitivity weights superimposed (coloured squares).
The regions with interference between $B^0 \rightarrow \bar{D}^{**0} K^0$ and $B^0 \rightarrow D^{**0} K^0$ colour 
suppressed processes show the greater sensitivity to $2\beta + \gamma$.
A particularly sensitive zone is at the intersection between $B^0 \rightarrow D^{-} K^{*+}$  and the colour 
suppressed $B^0 \rightarrow D^{**0} K^0$. In the same figure we can also notice that some sensitivity is also 
present because of the interference of the $D_{s,2}(2573)^{\pm}$ with the excited K resonances 
($K^{*}_0(1430)^{\pm}, K^{*}_2(1430)^{\pm}, K^{*}(1680)^{\pm}$).
As expected, we observe no sensitivity in the regions where there is only one path ($b \ra c$ or $b \ra u$) to reach the
$D K^0 \pi$ final state and a maximal sensitivity when there is an overlap between resonances due to both $b \ra c$ and $b \ra u$   
transitions. 

\section{Feasibility study}

\subsection{Dependence on the Dalitz model of the 2$\beta +\gamma$ error.}

\begin{figure}[!tbp]
\begin{center}
\epsfig{figure=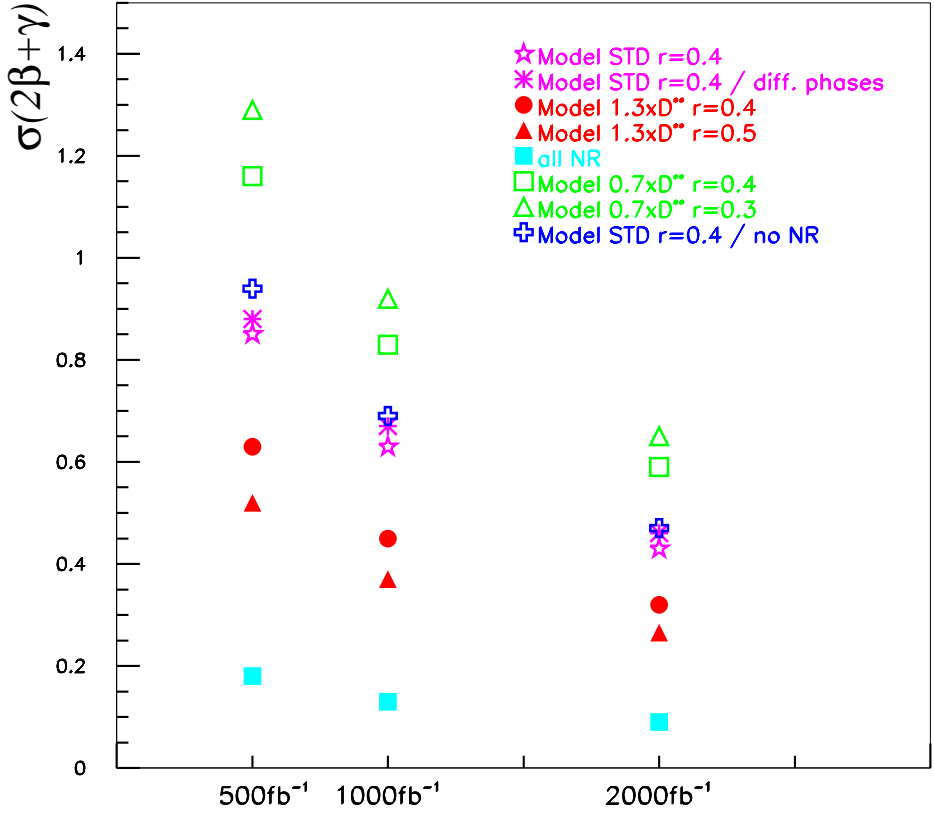,width=14.0cm}
\end{center}
\caption{\it{  Average absolute uncertainty obtained on the parameter $2\beta + \gamma$ as a function of the integrated luminosity for different models.
The magenta thick stars refers to the model in table \ref{tab:resonances} while the thin stars are obtained 
by varying the strong phases values; the red (dark) full circles and the green (clear) open square are obtained  
multiplying the $D^{**0}$ amplitudes by a factor 1.3 and 0.7 respectively. The red full (dark) and green (clear) open 
triangles refer to previous models considering $r$=0.5 and $r$=0.3 respectively.
Finally the blue crosses refers to the model in table \ref{tab:resonances} where the non-resonant component 
is set to zero.
The cyan (clear) squared are just given as reference and correspond to the simplified model presented 
in \cite{ref:ap} in which all $V_{ub}$ processes are considered through non-resonant amplitudes.}}
\label{fig:errormodel}
\end{figure}

The first study is to consider the dependence on the determination of 2$\beta + \gamma$ from the Dalitz structure of the
decay model. For this study we have used the time evolution equations defined in Eq. \ref{timelikeli} without including 
the effect of time resolution. Preliminary studies using a realistic resolution indicate an increase on 2$\beta +\gamma$ 
uncertainty of about 10$\%$. Instead, we have included realistic $\Bz \-- \Bzb $ tagging performances in 
terms of purities and efficiencies. The effect of the background is studied in the following section. 
For this study we just fit 2$\beta + \gamma$ fixing all the other parameters describing the Dalitz decay model.
The impact on the error on 2$\beta + \gamma$ by varying the decay model is studied as a function of the luminosity 
and shown in Figure \ref{fig:errormodel}.

\begin{table}
\begin{center}
\begin{tabular}{|c|c|c|}                                                    
\hline
 2$\beta+\gamma$ &  Parameter fitted           &   Configuration            \\ \hline
       0.63      &  $\twobg$         &    signal only             \\ \hline
       0.69      &  $\twobg$         &    signal only             \\
                 &  + $V_{cb}$ amp./phases     &    tagged only             \\ \hline
       0.65      &  $\twobg$         &    signal only             \\
                 &  + $V_{cb}$ amp./phases     &    tag + untag             \\ \hline
       0.79      &  $\twobg$         &    signal only             \\
                 &  + $V_{cb}$ amp./phases     &    tag + untag             \\
                 &  + $V_{ub}$ phases, $r$-fix &                            \\ \hline   
       0.84      &  $\twobg$         &    signal only             \\
                 &  + $V_{cb}$ amp./phases     &    tag + untag             \\
                 &  + $V_{ub}$ phases, amp.    &                            \\ \hline   
        0.82     &  $\twobg$         &    signal + 50$\%$ back.   \\ \hline
        0.85     &  $\twobg$         &    signal + 50$\%$ back.   \\  
                 &  + $V_{cb}$ amp./phases     &    tag + untag             \\ \hline
        1.00     &  $\twobg$         &    signal + 50$\%$ back.   \\
                 &  + $V_{cb}$ amp./phases     &    tag + untag             \\
                 &  + $V_{ub}$ phases, $r$-fix &                            \\ \hline   
     not conv.   &  $\twobg$         &    signal + 50$\%$ back.   \\
                 &  + $V_{cb}$ amp./phases     &    tag + untag             \\
                 &  + $V_{ub}$ phases, amp.    &                            \\ \hline
\end{tabular}
\end{center}
\caption{ \it {Average absolute uncertainty obtained on the parameter $2\beta + \gamma$ as a function of the fit configuration for an 
integrated luminosity of 1$ab^{-1}$. The configurations are explained in the text. See also Figure \ref{plot-error}. }}
\label{tab:results}
\end{table}

\begin{figure}[!tbp]
\begin{center}
\epsfig{figure=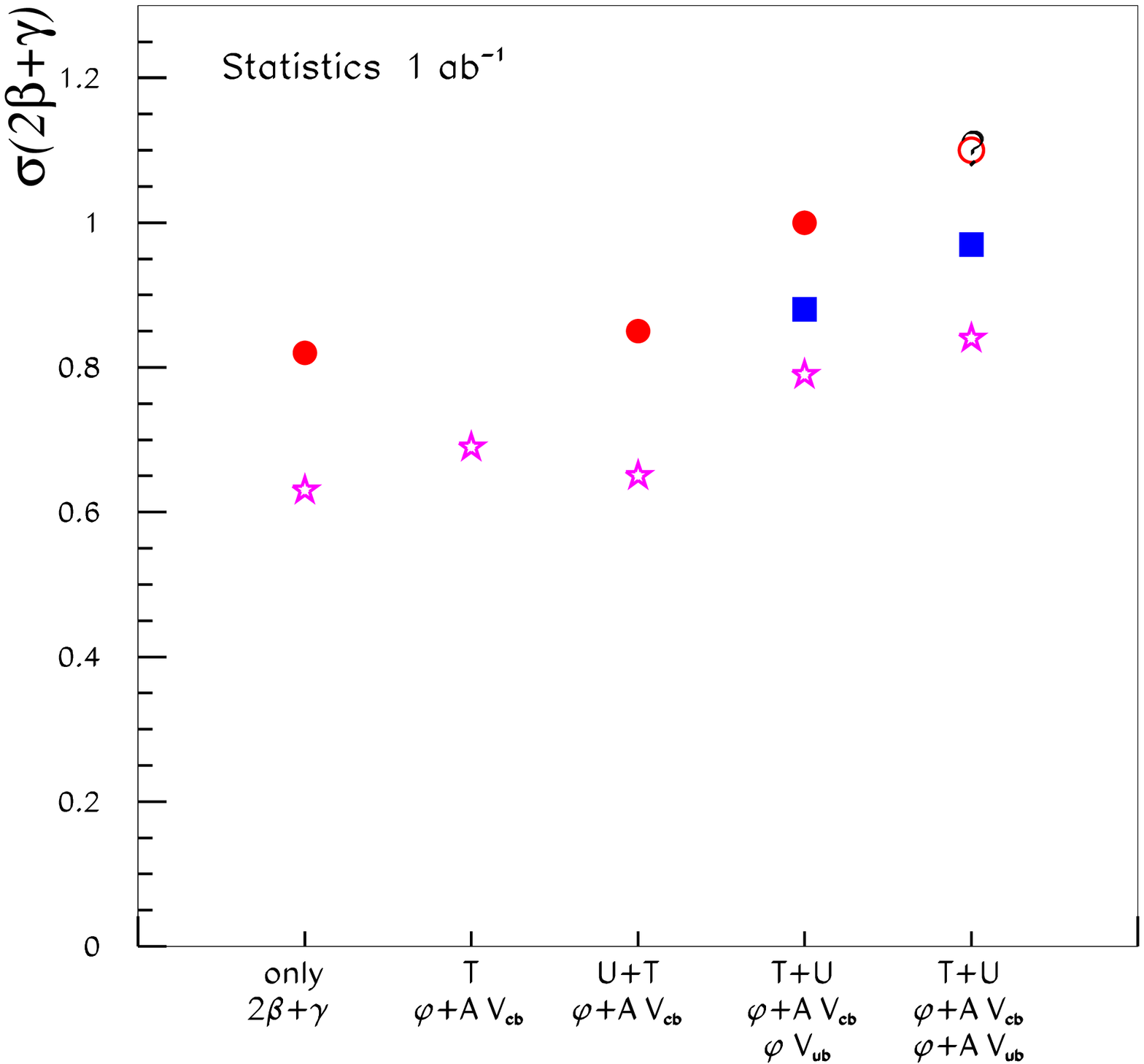,width=14.0cm}
\end{center}
\caption{ \it{ Average absolute uncertainty obtained on the parameter $\twobg$ as a function of the fit configuration.
The magenta thick stars refers to fit with no background while red dots and blue square correspond to fit
where the level of background is set at 50$\%$ and 30$\%$. The background has been simulated flat over the Dalitz plot.
The fit configuration are explained in the text.}}
\label{plot-error}
\end{figure}

From this study the first conclusion is that the present measurements \cite{ref:babarmh},\cite{ref:babarChinois} are 
not sufficient to fix the Dalitz model and give a precise indication on the error which can be obtained on 2$\beta +\gamma$.
 In fact, the precision on $\twobg$ strongly depends on the variation of the 
branching fractions of neutral $B$ into $D^{**}$ states within the measured errors. 
There is also some dependence on the presence or not of non resonant contributions and on the values of strong 
phases for the most relevant resonances. We also show the impact of the variation of 
$r=\frac{B^0 \rightarrow \bar{D}^{**0} K^0}{B^0 \rightarrow D^{**0} K^0}$ from 0.3 to 0.5.

The model using only a non resonant contribution for the $V_{ub}$ component is also given as a reference
since it was taken as example in the feasibility study presented in \cite{ref:ap}. The corresponding 
error on 2$\beta +\gamma$ is a factor 3-6 better with respect the one obtained using a set of realistic decay 
models. This model has not been further considered as it is contradiction with present data \cite{Aubert:2004at}.

\subsection{Model Independent fit}
An important feature of this method is the possibility to have a model independent determination of
$\twobg$ by fitting the parameters of the decay model.
We consider a sample corresponding to 1 \invab of collected statistics.
The reference error is obtained as previously by fitting only $\twobg$ and by fixing all the
other parameters. We try to progressively release the other parameters which characterise the decay model.

First we relax the  $D^{**}$ and the non-resonant $V_{cb}$ components. We managed to fit the 
three amplitudes and phases and the error on $\twobg$ increases by 10$\%$.
So far we have used only tagged events since they are the only one carrying information on $\twobg$. 
Nevertheless also untagged events could play an important role in fitting the decay model parameters. 
In practice we can add the following term to the likelihood given in Eq.\ref{timelikeli} :
\begin{equation}
P_{UNTAG}(B^{0} \rightarrow D^- K^0 \pi^+)={\frac{A_{c_k}^2+A_{u_k}^2}{2}}
\label{untaglikeli}
\end{equation}
The same fit is repeated with untagged events included in the sample. 
The error on $\twobg$ decreases by about 10$\%$.
In the following untagged events are used. \\ 
The following test is to leave free the strong phases of the $D^{**}$ and the non-resonant $V_{ub}$ 
components. In this case the error on $\twobg$ increases by about 20$\%$. An extra increase of
10$\%$ is obtained if  the amplitudes of those components are further left free to vary in the fit.

One can also try to estimate the effect of the presence of the background in the Dalitz plot. For simplicity we take
a flat background over the Dalitz plot, being aware that reality could be different. 
We  assume that the discrimination between signal and background events is performed using additional 
variables ({\it e. g.} the reconstructed $B$ mass) and  we fix the signal over background ratio 
in the CP violation fit. We run our simulation 
by considering a signal over background (S/B) ratio of 30$\%$ or 50$\%$. The result is that the error on 
$\twobg$ increases by 25$\%$ and 50$\%$, respectively. Furthermore if S/B is equal to 50$\%$
it seems difficult to perform a fit where all $V_{cb}$ and $V_{ub}$ $D^{**}$ and the non-resonant components
are left free in the fit. \\
As by-product of this analysis, the Dalitz model can be fitted from the data if the level of background 
is not too large. The $V_{cb}$ $D^{**}$ related amplitudes and phases can be fitted with precisions 
lying between 20-30$\%$ . Furthermore both the phases and the amplitudes for $D^{**}$ from $V_{ub}$ processes 
are determined with errors which can go up to 50$\%$.
In addition the excited K resonnaces ($K^{*}_0(1430)^{\pm}, K^{*}_2(1430)^{\pm}, K^{*}(1680)^{\pm}$) 
which concerns only $V_{cb}$ processes can be precisely determined from data with relative uncertainties 
better than 25$\%$.



\section {Conclusions and perspectives}

We have performed a full feasibility study for measuring $\twobg$ 
with time dependent Dalitz analysis using $D^- K^0 \pi^+$ final states.
For this study a realistic decay model, based on the available experimental 
information has been elaborated.

We conclude that the error on $\twobg$ strongly depends on the decay 
model and that the currently available experimental information is not enough for a 
precise estimate. By varying the different parameters in the currently allowed range, 
the error on $\twobg$ can vary within a factor of about 2.5. Only a complete 
Dalitz analysis could tell us where we stand. This is possible on data. \\
In fact we have 
shown that if the level of background is not too large ($<50 \%$) the full decay model can 
be obtained by fitting the Dalitz plot using both tagged and untagged events.

All the amplitudes and strong phases can be fitted and, in case of the Dalitz decay model given 
in Table \ref{tab:resonances}, the precision on $\twobg$ could be of about 50$\%$ 
with 500 \invfb and should not be limited by the systematical uncertainties.
This implies that the precision on 2$\beta + \gamma$ could be reduced at few percent level
in case this analysis is performed at a high luminosity B factory \cite{ref:superb}.

\section{Acknowledgement}
We would like to thanks Anne-Marie Lutz, Patrick Roudeau and Riccardo Faccini for 
useful suggestions and for the careful reading of this paper.

\end{document}